\documentclass[11pt, letterpaper]{article}

\usepackage{amsmath,amssymb,amsfonts}
\usepackage{graphicx}
\usepackage{float}
\usepackage{colortbl}
\usepackage{booktabs}
\usepackage{multirow}
\usepackage{tabularx}
\usepackage{hyperref}
\usepackage[numbers]{natbib}
\usepackage{geometry}
\usepackage{enumitem}
\usepackage{array}
\usepackage[table]{xcolor}

\newcolumntype{C}{>{\centering\arraybackslash}X}
\newcolumntype{L}{>{\raggedright\arraybackslash}X}

\hypersetup{
  colorlinks=true,
  linkcolor=black,
  citecolor=black,
  urlcolor=blue,
  hidelinks
}

\title{Measuring Complexity at the Requirements Stage: Spectral Metrics as Development Effort Predictors}

\author{
Maximilian Vierlboeck$^{1,*}$, Antonio Pugliese$^{1}$, Roshanak Rose Nilchiani$^{1}$, \\ Paul T. Grogan$^{2}$, Rashika Sugganahalli Natesh Babu$^{1}$ \\[6pt]
\small $^{1}$ School of Systems and Enterprises, Stevens Institute of Technology, Hoboken, NJ 07030, USA \\
\small $^{2}$ School of Computing and Augmented Intelligence, Arizona State University, Tempe, AZ 85281, USA \\[4pt]
}

\date{}

\begin{document}
\maketitle

\begin{abstract}
Complexity in engineered systems presents one of the most persistent challenges in modern development since it is driving cost overruns, schedule delays, and outright project failures. Yet while architectural complexity has been studied, the structural complexity embedded within requirements specifications remains poorly understood and inadequately quantified. This gap is consequential: requirements fundamentally drive system design, and complexity introduced at this stage propagates through architecture, implementation, and integration. To address this gap, we build on Natural Language Processing methods that extract structural networks from textual requirements. Using these extracted structures, we conduct a controlled experiment employing molecular integration tasks as structurally isomorphic proxies for requirements integration---leveraging the topological equivalence between molecular graphs and requirement networks while eliminating confounding factors such as domain expertise and semantic ambiguity. Our results demonstrate that spectral measures predict integration effort with correlations exceeding 0.95, while structural metrics achieve correlations above 0.89. Notably, density-based metrics show no significant predictive validity. These findings indicate that eigenvalue-derived measures capture cognitive and effort dimensions that simpler connectivity metrics cannot. As a result, this research bridges a critical methodological gap between architectural complexity analysis and requirements engineering practice, providing a validated foundation for applying these metrics to requirements engineering, where similar structural complexity patterns may predict integration effort.
\end{abstract}

\noindent\textbf{Keywords:} systems engineering; complexity; requirements engineering; quantitative metrics; spectral graph theory; graph energy

\bigskip

\section{Introduction}
\label{sec:introduction}

Modern engineered systems are growing increasingly complex. The integration of diverse subsystems, emergent behaviors, and interdependent networks introduces significant challenges in system design, development, and maintenance \cite{pugliese2018development}. These challenges affect both the engineers developing these systems and those operating them. As systems become more intricate, understanding and managing their complexity is critical to ensuring their reliability, efficiency, and overall success, as the consequences could otherwise be schedule overruns, defects \cite{sturtevant2013system}, development stops, and even complete product failures~\cite{hussain2016role, mukherjee2008understanding}. Requirements, as the foundational specifications guiding system development, play a crucial role in defining system functionality and performance \cite{Lindemann1998, boznak1994doing}. However, the inherent complexity of requirements---driven by interdependencies, ambiguities, and structural intricacies---can significantly impact the development process, leading to unforeseen risks, cost overruns, and schedule delays \cite{6549917, sinha2016empirical, azmat2023analyzing}. Despite this recognized impact, no empirically validated metrics exist for quantifying the structural complexity of requirements in a way that could predict or gauge development effort---a gap that leaves practitioners without objective tools for early-stage complexity assessment.

In practice, this gap is consequential. Systems engineers routinely assess requirements for completeness, consistency, and traceability, yet these evaluations remain largely qualitative when it comes to structural complexity \cite{vierlboeck2023re, salado2014concept}. When a specification grows to hundreds or thousands of interconnected requirements, practitioners lack scalable methods to determine which clusters of requirements will demand disproportionate integration effort or where hidden interdependencies may propagate risk across subsystems. As a result, complexity-driven problems are typically discovered late---during integration and test---when the cost of correction is highest \cite{Lindemann1998}. Recent work has demonstrated that structural complexity metrics applied to system models can predict development effort and cost~\cite{bhatnager2025measuring}, yet equivalent validated measures for requirements specifications remain unavailable. An objective, computable measure of structural complexity at the requirements stage would enable earlier intervention, more informed resource allocation, and reduced downstream rework.

Effectively managing complexity requires rigorous, quantifiable metrics that provide insights into system structure and behavior. Traditional approaches to complexity assessment primarily focus on system architecture, using metrics that analyze connectivity, modularity, and component interactions \cite{gutman2006laplacian, nikiforov2007energy, sinha2016empirical, 6549917}. While these methods offer valuable perspectives on architectural complexity, they do not consider the structural complexity embedded within system requirements. Requirements engineering (RE), predominantly expressed in natural language, introduces additional layers of complexity that are difficult to quantify using conventional architectural metrics \cite{salado2014concept}. Prior work has established that natural language processing techniques can extract structural networks from requirements text \cite{SysCon2022, vierlboeck2025natural}, and that spectral metrics such as Graph Energy, which was originally developed from chemistry approaches, can quantify complexity based on the spectrum of a graph \cite{pugliese2018development}. Yet whether these spectral metrics actually predict human performance outcomes remains empirically untested.

Spectral metrics are particularly promising for this purpose because they encode global structural information that simpler measures miss. Density, for instance, captures the ratio of actual to possible connections but is insensitive to how those connections are arranged topologically---two networks with identical density can present vastly different integration challenges depending on clustering, path structure, and hierarchical organization. Graph Energy and Laplacian Graph Energy, by contrast, are derived from the full eigenvalue spectrum of a graph's adjacency and Laplacian matrices, respectively \cite{gutman2006laplacian, nikiforov2007energy}. The eigenvalue distribution encodes information about paths, patterns, and structural regularity, capturing dimensions of complexity that directly affect how engineers comprehend and integrate interconnected components \cite{pugliese2019developing}. This theoretical advantage motivates their selection as candidate predictors of integration effort.

This challenge takes on new urgency as large language models (LLMs) increasingly enter requirements engineering workflows \cite{mehraj2024tertiary, tikayat2023agile}, creating a need for validated metrics to assess structural complexity of specifications, whether human-written or AI-supported.

This paper addresses that gap through empirical validation. We present a controlled experimental case study that tests whether complexity metrics predict integration effort in requirement-like structures. By leveraging mathematical constructs such as adjacency matrices, Graph Energy, and Laplacian Energy, we demonstrate that spectral complexity metrics correlate strongly with task completion time in structurally isomorphic integration tasks, suggesting that measured structural complexity may predict effort in analogous requirements contexts. Our approach builds upon prior research in system complexity analysis \cite{pugliese2019developing, sinha2016empirical} and extends its application to requirements \cite{vierlboeck2025natural, SysCon2022, salado2014concept}, addressing a critical need in systems engineering \cite{pugliese2018development}.

To enable controlled investigation, we employ an isomorphic task design using molecular integration tasks. The study draws an analogy between molecular structures and requirements networks \cite{vierlboeck2025natural, vierlboeck2023re}, leveraging the structural correspondence that both can be represented as graphs with nodes and edges. Through empirical analysis of complexity metrics across twenty-three participants, we reveal that structural metrics (Graph Energy, Laplacian Graph Energy) demonstrate strong predictive validity ($r > 0.89$), while density-based metrics show no significant correlation with integration effort. These findings provide actionable guidance for metric selection in requirements engineering practice.

This work makes four primary contributions: (i)~a methodological framework that leverages structural isomorphism between molecular graphs and requirement networks to enable controlled empirical investigation of complexity metrics independent of domain expertise and semantic confounds; (ii)~empirical validation demonstrating that spectral metrics---Graph Energy and Laplacian Graph Energy---predict integration effort with correlations exceeding 0.95, with structural metrics such as Integration Load and Cyclomatic Complexity also achieving correlations above 0.89, while density-based metrics show no predictive validity; (iii)~a preliminary foundation for extending complexity assessment to requirements engineering through NLP-based structural extraction; and (iv)~a framework for proactive complexity management, including integration into emerging AI and LLM-supported workflows where objective structural quality gates are needed.

The remainder of this paper is structured as follows: Section~\ref{sec:background} reviews relevant literature on complexity metrics and their applications in engineered systems. Section~\ref{sec:methodology-setup} details the methodology and experimental setup. Section~\ref{sec:results} presents the case study results, followed by a discussion in Section~\ref{sec:discussion} on their implications for systems engineering, including applications to LLM-assisted requirements development. Finally, Section~\ref{sec:conclusion} concludes with limitations and recommendations for future research and practice.

\section{Background}
\label{sec:background}

\subsection{Complexity and Its Metrics in Engineered Systems}

The complexity of engineered systems is a defining characteristic that influences their design, integration, and long-term evolution. As modern systems become increasingly intricate---spanning cyber-physical, aerospace, and socio-technical domains---understanding and managing complexity has emerged as a critical challenge in systems engineering \cite{sheard20107, vandergriff2007system}. The integration of diverse subsystems, emergent behaviors, and interdependent networks introduces significant challenges that affect both the engineers working on development and the users interacting with these systems. As systems become more complex, their reliability, efficiency, and overall success depend on rigorous complexity management, as the consequences of inadequate control can manifest as schedule overruns, development stoppages, and even complete product failures \cite{hussain2016role, mukherjee2008understanding}.

Early complexity studies, such as those by Weaver \cite{weaver1948}, classified problems into simple, disorganized complexity, and organized complexity. Modern complexity science \cite{richardson2000complexity} has extended this classification, distinguishing between structural complexity \cite{sinha2014structural}, which is driven by system architecture and interconnections, and behavioral complexity, which emerges from dynamic interactions \cite{sinha2016empirical}. Recent research has further refined these distinctions, exploring how emergent behavior in complex systems can be detected and predicted through machine learning approaches that combine live and post-mortem analysis techniques \cite{dahia2024emergent}. These developments highlight the evolving nature of complexity theory, where computational methods increasingly complement traditional analytical frameworks.

\subsubsection{Defining and Measuring Complexity in Engineered Systems}

Several methodologies have been developed to quantify complexity. Among the most widely used are information-theoretic metrics, graph-theoretic approaches, and structural complexity models that have evolved significantly over the past two decades:

\begin{itemize}
    \item \textit{Information-Theoretic Metrics:} Shannon's entropy \cite{shannon1948mathematical} remains foundational in complexity measurement, quantifying system uncertainty. Expanding on this, Gell-Mann and Lloyd \cite{gell1996information} introduced effective complexity, distinguishing structured information from randomness---a distinction critical for understanding organized versus disorganized complexity in engineered systems \cite{snowden2007leader}. Information-theoretic approaches have found renewed application in recent years, with spectral entropy being utilized to assess the complexity of requirements specifications and system architectures \cite{6549917}. These metrics provide a mathematical basis for quantifying the information content and structural organization inherent in system descriptions.

    \item \textit{Graph-Theoretic Approaches:} Structural metrics leverage the eigenvalue distributions of adjacency matrices to evaluate system complexity through network representations \cite{herrera2020multi}. Gutman and Zhou's \cite{gutman2006laplacian} Laplacian Energy metric and Nikiforov's \cite{nikiforov2007energy} Graph Energy metric have been particularly influential in quantifying topological complexity. These foundational works established the basis for spectral analysis of system architectures, enabling the assessment of connectivity patterns and interaction structures \cite{strogatz2001exploring}. Building upon these classical metrics, Pugliese and Nilchiani \cite{pugliese2019developing} developed a generalized framework for spectral structural complexity metrics that unified distinct measures through variations in mathematical functions, coefficients, and matrix representations. This framework was successfully applied to cyber-physical systems, demonstrating the predictive power of Graph Energy in assessing system resilience and identifying tipping points where systems transition from stable to unstable states \cite{edwards2024tipping, edwards2024resilience}. Recent advances have extended spectral complexity analysis to directed graphs \cite{mezic2019spectral}, though the current work focuses on undirected representations to maintain consistency with established architectural complexity frameworks \cite{pugliese2019developing, sinha2016empirical}.

    \item \textit{Structural Complexity Models:} Sinha and de Weck \cite{sinha2016empirical} validated a structural complexity metric that integrated graph-theoretic principles with system architecture considerations. That model, widely applied in cyber-physical systems, captures connectivity, modularity, and system organization. That seminal work established empirical validation methodologies that linked complexity metrics to observable system properties and development challenges. Lopez and Thomas \cite{lopez2022metric} extended that framework to space systems modeled in SysML, adapting the adjacency-matrix-based complexity metric for hierarchical decomposition across multiple system levels. Subsequent research extended these structural approaches to address multiple design constraints \cite{sinha2020design}, Pareto-optimization of complexity versus modularity trade-offs \cite{sinha2018pareto}, and integrative complexity measures that provided alternative perspectives on system decomposition~\cite{sinha2018integrative}. Recent work has further explored how modularity and interdependence in cyber-physical systems can be analyzed using bio-inspired graph modeling approaches that draw from ecological systems theory \cite{bioinspired2024cps}, demonstrating the cross-pollination of complexity concepts across disciplinary boundaries. Similarly, Bhatnager et al.\ \cite{bhatnager2025measuring} proposed structural complexity metrics for SysML models, demonstrating that quantifying model complexity can predict development effort and inform engineering decisions.

    Beyond spectral approaches, graph-theoretic metrics such as McCabe's Cyclomatic Complexity \cite{mccabe1976complexity} have been widely applied in software engineering to measure logical complexity based on the number of linearly independent paths through a program's control flow graph. While originally developed for software code, such structural metrics provide complementary perspectives to spectral analysis and have been adapted for system-level complexity assessment.
\end{itemize}

\subsubsection{Implications for Systems Engineering}

The management of complexity is directly tied to system risks, cost overruns, and integration challenges. Studies have linked increased structural complexity with reduced predictability in large-scale aerospace and software projects, necessitating proactive complexity quantification early in development \cite{collopy2011value, nilchiani2017systems, azmat2023analyzing}. Nilchiani and Pugliese \cite{nilchiani2017systems} established strong correlations between quantitative complexity levels of acquisition programs and the likelihood of technical failures, demonstrating that complexity-based risk assessment can inform resource allocation and mitigation strategies. Understanding complexity at multiple levels has become critical for ensuring robust system performance across increasingly interconnected and interdependent systems.

High structural complexity often correlates with increased integration challenges, making it essential to identify early indicators of complexity before they escalate into major engineering risks. Prior research suggests that spectral complexity metrics, particularly Graph Energy and Laplacian Energy, may serve as predictors of integration time and cognitive load \cite{sinha2016empirical, pugliese2019developing}. This relationship is investigated by the presented study through controlled experimentation. Hence, the findings provide quantitative evidence that complexity management must begin at the requirements stage, before architectural decisions constrain design flexibility.

Emerging tools and methodologies are enhancing traditional complexity assessment methods. Machine learning-driven complexity analysis enables pattern recognition in large-scale system models, identifying complexity hotspots and predicting emergent behaviors that may not be apparent through conventional analytical approaches \cite{dahia2024emergent}. Automated requirements processing can enhance traditional complexity assessment methods, providing decision-makers with real-time insights into evolving system complexity \cite{salado2014concept}.

Moreover, complexity considerations extend beyond technical challenges to organizational and life-cycle management aspects. Managing system complexity effectively requires a holistic approach that integrates engineering principles with project management methodologies, ensuring that evolving system requirements remain aligned with stakeholder objectives. With this foundation in complexity metrics and their implications for systems engineering, the following section introduces the spectral metrics employed in this research.

\subsection{Spectral Structural Complexity Metrics}

To quantify spectral complexity, we employ three weighted matrix representations of the system graph: the adjacency matrix $\bar{A}(u, v)$, the Laplacian matrix $\bar{L}(u, v)$, and the normalized Laplacian matrix $\bar{\mathcal{L}}(u, v)$. The weighting scheme encodes both node and edge attributes, where weights correspond to the complexity contributions of individual components and their interfaces as formalized by Sinha \cite{sinha2014structural, sinha2012structural}. The mathematical foundations for these matrix formulations follow Chung \cite{chung1997spectral} and Spielman \cite{spielman2007spectral}.

\subsubsection{Graph Energy}

Gutman \cite{gutman2001energy, gutman2011hyperenergetic} introduced Graph Energy as the sum of absolute eigenvalues of the adjacency matrix:
\begin{equation}
E_A(G) = \sum_{i=1}^n |\lambda_i|
\end{equation}
This metric satisfies several key properties \cite{gutman2006laplacian}: (1) non-negativity, with $E_A(G) = 0$ only when no edges exist ($m = 0$); (2) additivity over disconnected components, such that $E_A(G) = E_A(G_1) + E_A(G_2)$ for disjoint subgraphs $G_1$ and $G_2$; and (3) invariance to isolated vertices, where $E_A(G) = E_A(G_1)$ if all components beyond $G_1$ are singletons.

For weighted graphs, we extend this definition using the weighted adjacency matrix:
\begin{equation}
E_{\bar{A}}(G) = \sum_{i=1}^n |\bar{\lambda_i}|
\end{equation}
where $\bar{\lambda_i}$ denotes the $i$th eigenvalue of $\bar{A}$.

To enable meaningful comparison across graphs of varying size, this formula incorporates a normalization coefficient $\gamma = 1/n$. This adjustment accounts for the fact that the number of components in a system representation reflects architectural decomposition choices rather than inherent system complexity. Applying this coefficient yields:
\begin{equation}
E_{\bar{A},n}(G) = \frac{1}{n} \sum_{i=1}^n |\bar{\lambda_i}|
\end{equation}
where $\bar{\lambda_i}$ denotes the eigenvalues of $\bar{A}$. This normalization convention aligns with approaches used in natural connectivity \cite{wu2010natural} and structural complexity measurement \cite{sinha2012structural}.

\subsubsection{Laplacian Graph Energy}

Gutman and Zhou \cite{gutman2006laplacian} extended Graph Energy to the Laplacian matrix representation:
\begin{equation}
E_L(G) = \sum_{i=1}^n \Big|\mu_i - \frac{2m}{n} \Big|
\end{equation}
where $\mu_i$ denotes the $i$th eigenvalue of the Laplacian matrix, $n$ is the number of nodes, and $m$ is the number of edges.

\subsubsection{Normalized Laplacian Graph Energy}

Cavers et al. \cite{cavers2010normalized} proposed an analogous metric using the normalized Laplacian:
\begin{equation}
E_{\mathcal L}(G) = \sum_{i=1}^n | \nu_i - 1 |
\end{equation}
where $\nu_i$ represents the eigenvalues of the normalized Laplacian matrix $\mathcal{L}$.

\subsubsection{Natural Connectivity}

Wu et al. \cite{wu2010natural} developed natural connectivity as a robustness-oriented metric:
\begin{equation}
N_{A,n}(G) = \ln \Biggl( \frac{1}{n} \sum^n_{i=1} e^{\lambda_i} \Biggr)
\end{equation}
where $\lambda_i$ are the eigenvalues of the adjacency matrix. This formulation incorporates the $\gamma = 1/n$ normalization coefficient discussed previously.

\subsection{Complexity in Requirements Engineering}
\label{subsec:complexity-in-re}

A significant gap exists between how we measure architectural complexity and how we assess complexity in requirements. Most work on structural complexity focuses on system architecture---analyzing component interactions, network topology, and modularity---once the architecture is largely defined \cite{sinha2016empirical, sinha2018integrative}. Existing architectural metrics \cite{sinha2020design, salado2014concept, sinha2018pareto}, while effective in structural analysis, do not work with requirements and their structure, as they rely on further developed system components and sub-components. Yet, requirements engineering, which fundamentally drives architectural decisions, has few tools for assessing structural complexity. This matters because failing to manage requirements complexity early has substantial consequences: difficulty predicting development challenges, inability to identify problematic requirement structures before they cascade through design and implementation, and inability to leverage spectral complexity approaches that have proven effective in architectural analysis.

Beyond architectural considerations, requirements themselves contribute to development complexity by adding to the load and or effort \cite{mccabe1976complexity, halstead1977elements}. Establishing causal links between various complexity metrics remains challenging, yet each metric captures a distinct facet of the overall system and process complexity \cite{salado2014concept}. Furthermore, although system architectures emerge early in development, their structural detail increases progressively over time. In initial phases, architectures typically lack sufficient interconnections to support meaningful complexity quantification and comparison \cite{vierlboeck2025natural}.

\subsubsection{Existing Complexity Metrics in Requirements Engineering}

Traditional complexity metrics in RE predominantly focus on syntactic and semantic properties but lack the capacity for comprehensive structural analysis. Three categories of approaches have emerged.

\textit{Text-Based Complexity Metrics:} Conventional RE complexity metrics evaluate syntactic properties such as requirement length, readability scores, and linguistic ambiguity detection \cite{Berry2004, berry2004user, kamsties2001detecting}. While these metrics provide useful insights, they do not encapsulate the deeper structural dependencies within requirement sets.

\textit{Network-Based Metrics:} Some methodologies model requirements as networks, analyzing graph properties such as connectivity and centrality \cite{hein2022reasoning, wang2014network}. While these approaches offer valuable perspectives on interdependencies, they often do not fully leverage spectral graph properties, limiting their ability to quantify deeper structural complexities.

\textit{Structural and Spectral Metrics:} Emerging approaches employ spectral analysis to assess requirement complexity by incorporating graph-theoretic principles \cite{gutman2006laplacian, sinha2014structural}. However, these metrics have been developed primarily for system architectures, and their application to requirements specifications prior to architectural definition remains underexplored.

\subsubsection{Bridging the Gap with Spectral Metrics}

Addressing this gap early in development is critical. When requirements drive architectural decisions, quantifying complexity at the requirement level becomes essential to identify issues before costs escalate \cite{arena2008cost}. The earlier problematic requirements, circular dependencies, excessive coupling, and ambiguous specifications can be identified, the better positioned teams are to address them when design flexibility is still available and intervention costs are manageable.

Prior work by some of the authors \cite{vierlboeck2025natural} demonstrated that spectral complexity metrics could be applied directly to requirements using Natural Language Processing. Rather than waiting for formal architecture development, that method extracted structural information from textual requirements by identifying three key layers: entities (components and actors mentioned), relationships (how requirements reference each other), and hierarchical organization (the decomposition structure) \cite{SysCon2022}. From these extracted layers, weighted adjacency matrices were constructed and spectral complexity measures computed---Graph Energy, Laplacian Energy, and entropy. Validation showed this approach achieved high precision in structural detection and effectively identified problematic patterns like circular dependencies and complexity hotspots.

Spectral complexity metrics present a promising avenue to bridge the gap between architectural and requirements complexity by:
\begin{itemize}
    \item Capturing latent structural dependencies within requirements;
    \item Identifying clusters of interdependent requirements;
    \item Quantifying complexity beyond text-based heuristics, offering a rigorous mathematical framework for analyzing requirement interactions.
\end{itemize}

By integrating spectral complexity metrics into RE, this research enhances structural analysis methodologies traditionally reserved for later-stage system architecture assessment, enabling proactive complexity management at the inception of system development.

Integrating complexity metrics into the initial phases of system development is paramount to mitigating downstream design risks, cost overruns, and schedule delays. Requirements fundamentally drive system architecture, and quantifying complexity at the requirement level facilitates proactive complexity management prior to the formalization of system structures \cite{vierlboeck2025natural}. A robust complexity quantification framework for textual requirements enhances clarity, reduces ambiguity, and improves traceability throughout the engineering life cycle.

Building upon the NLP-based extraction methodology \cite{vierlboeck2025natural, SysCon2022}, the present research extends spectral complexity assessment by empirically validating the relationship between measured complexity and integration effort. This validation addresses a critical gap: while prior work established that complexity could be measured from requirements, whether that measured complexity predicts development outcomes remained untested. This paper addresses this gap through controlled experimentation, bridging the theoretical and practical dimensions of requirements engineering.

\section{Methodology and Setup}
\label{sec:methodology-setup}

The methodology is presented in three stages. First, we establish how spectral complexity metrics developed for system architectures can be applied to requirements by exploiting the structural isomorphism between requirement networks and molecular graphs. Second, we describe a controlled experiment in which participants assemble molecular structures of varying complexity, with completion time serving as a proxy for the integration effort. Third, we define the specific metrics computed on each task and the statistical models used to evaluate their predictive validity.

\subsection{Extending Metrics to Requirements Engineering}
\label{subsec:extending-metrics}

To address the gap between architectural and requirements complexity outlined in Section~\ref{subsec:complexity-in-re}, in prior work, the authors developed a methodology that extracts latent structural information from natural language requirements and maps it to a graph-theoretic representation \cite{vierlboeck2025natural, SysCon2022}.

The process begins with an NLP-based parsing pipeline that identifies three distinct structural layers within a requirement specification \cite{vierlboeck2025natural}:

\begin{enumerate}
    \item Hierarchy Layer: The parent--child decomposition of requirements.
    \item Requirement Layer: The direct dependencies and cross-references between requirement statements.
    \item Entity Layer: The interaction of physical components and actors defined within the text, revealing implicit functional couplings.
\end{enumerate}

This extraction methodology achieved over 99 percent precision in structural detection when validated against expert-annotated requirements \cite{vierlboeck2023re}.

\begin{figure}[H]
    \centering
    \includegraphics[width=0.8\linewidth]{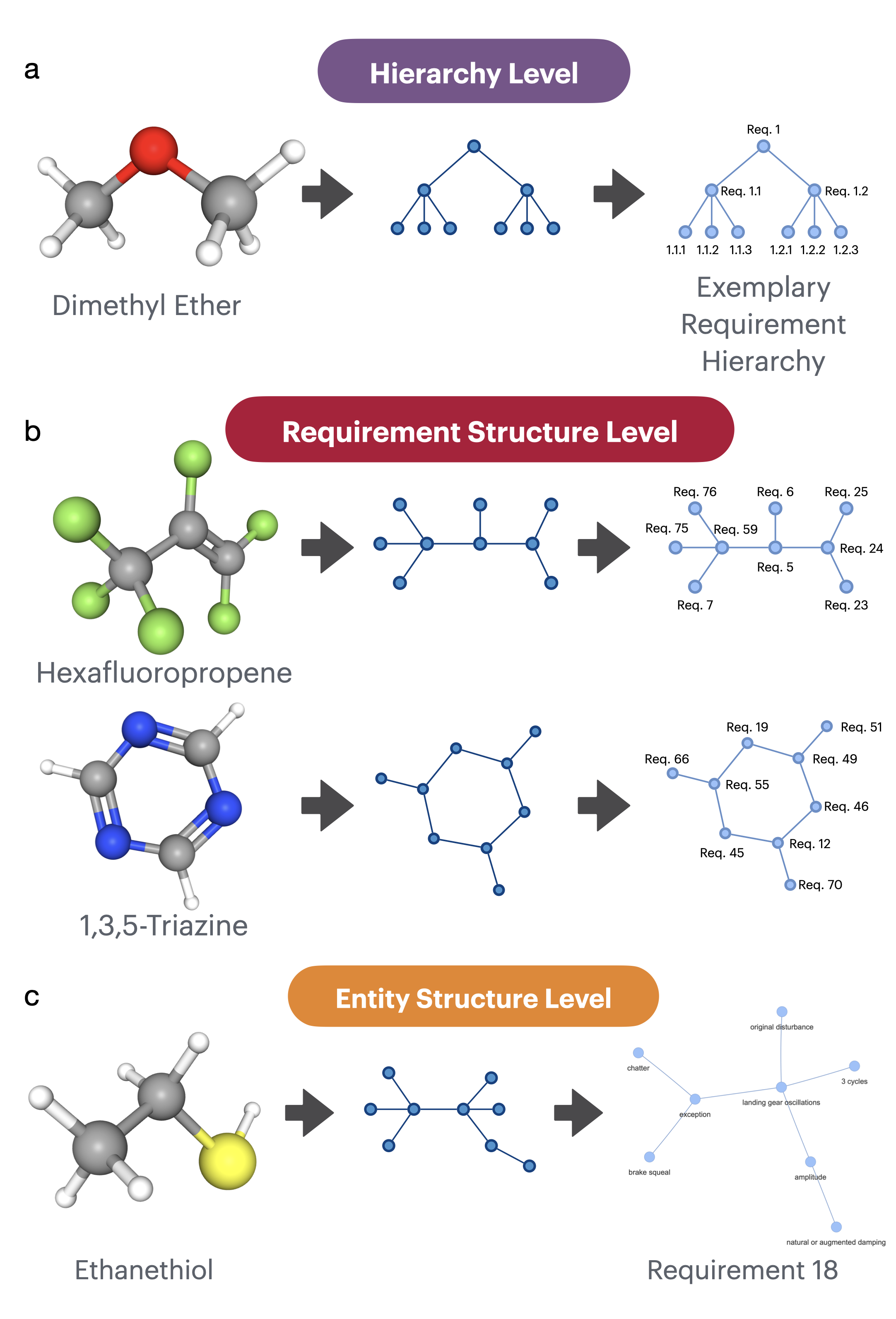}
    \caption{Structural isomorphism between molecular assemblies and requirement specifications. (a) Hierarchical decomposition of requirements mirrors branched molecular structures. (b) Interdependent requirement clusters exhibit cyclic topologies similar to ring-based molecules (e.g., hexafluoropropene, and 1,3,5-triazine). (c) Entity-level fragments from NLP extraction (e.g., Requirement 18) map to specific molecular signatures (e.g., ethanethiol), validating the use of molecular integration tasks as a proxy for requirement complexity.}
    \label{fig:requirementsanalogy}
\end{figure}

Once extracted, these layers form a network of nodes and edges that exhibits striking topological similarities to molecular structures. As illustrated in Figure~\ref{fig:requirementsanalogy}(a), the structural decomposition of requirements mirrors the hierarchical organization of chemical compounds. For instance, Figure~\ref{fig:requirementsanalogy}(a) demonstrates how the breakdown of a requirement into sub-requirements ($1 \rightarrow 1.1 \rightarrow 1.1.1$) maps directly to the branching structure of a dimethyl ether molecule.

Similarly, Figure~\ref{fig:requirementsanalogy}(b) shows how dense clusters of interdependent requirements, specifically from the Skyzer UAV landing gear specification \cite{vierlboeck2023re, blackburn2018skyzer}, topologically resemble the cyclic structure of ring-based molecules like 1,3,5-triazine. This recurring pattern of cyclic dependencies in requirements often indicates high coupling and integration difficulty, much like ring structures in chemistry dictate stability and reactivity. At the most granular level, Figure~\ref{fig:requirementsanalogy}(c) reveals that individual requirement fragments, such as Requirement 18 from the experimental dataset, share an almost identical graph signature with ethanethiol~\cite{vierlboeck2023re}.

This structural isomorphism validates the use of molecular integration as a proxy for requirement integration. Just as increasing the number of atoms and bond types in a molecule increases the cognitive load required to assemble it, increasing the number of entities and dependencies in a requirement specification increases the cognitive effort required for engineers to comprehend and implement the system \cite{pugliese2018development}. By quantifying the spectral properties of these graphs (via the adjacency and Laplacian matrices described below), we can predict the integration effort required for the system before a physical architecture is ever built.

\subsection{Metrics Framework}
\label{subsec:metrics-framework}

The generalized formula for a spectral structural complexity metric was introduced in \cite{pugliese2019developing} as
\begin{equation}
\label{eq:generalized-complexity}
C(S) = f \Big(\gamma \sum^n_{i=1} g \Big(\lambda_i(M) - \frac{tr(M)}{n} \Big) \Big)
\end{equation}
where the functions $f$ and $g$, the coefficient $\gamma$, and the matrix $M$ can be adjusted to obtain a series of metrics.
The values used for these variables are $(f(x), g(y)) = [(x, |y|),  (\text{ln}(x), e^y)]$, $\gamma = [1, 1/n]$, and $M = [\bar{A}, \bar{L}, \bar{\mathcal L}]$. In this study, all metrics reported in the results use the $\gamma = 1/n$ normalization unless otherwise noted, ensuring comparability across tasks with different numbers of components.

\subsubsection{Component, Interface, and Topological Complexity}

The spectral metrics address three distinct contributions to structural complexity~\cite{pugliese2018development, sinha2014structural}:
\begin{enumerate}
    \item \textit{Component complexity} ($\alpha_i$): The inherent complexity of individual elements within the system, determined by their internal structure and properties.
    \item \textit{Interface complexity} ($\beta_{ij}$): The complexity introduced by connections between components, modeled as $\beta_{ij} = \sqrt{\alpha_i \alpha_j}$, reflecting that interfaces between complex components are themselves more complex.
    \item \textit{Topological complexity}: The complexity arising from the overall arrangement and connectivity pattern of the system, captured through the matrix representation and its spectral properties.
\end{enumerate}

As noted by Pugliese \cite{pugliese2018development}, the hypothesis underlying these metrics is that ``a metric defined as a function of the spectrum of a matrix representation of the system of interest, where the matrix is weighted according to the complexity contributions of nodes and edges, is a monotonically increasing function of the assembly time of the system.'' Graph Energy and Laplacian Graph Energy primarily capture topological complexity through eigenvalue analysis, while composite metrics such as Integration Load combine all three contributions.

\subsubsection{Metrics Evaluated}

For this study, metrics were computed at two levels of analysis: the \textit{molecule level} (complexity of individual components before integration) and the \textit{integration level} (complexity introduced when combining components). This dual-level approach enabled the assessment of both inherent component complexity and emergent integration complexity.

\medskip
\textit{Spectral Metrics:}
\begin{itemize}
    \item \textit{Graph Energy (GE):} Defined by Gutman \cite{gutman2001energy} as the sum of absolute eigenvalues of the adjacency matrix, Graph Energy captures global structural properties including path lengths, clustering patterns, and structural regularity \cite{gutman2006laplacian}. Originally derived from H\"{u}ckel's \cite{huckel1931quantentheoretische} molecular orbital theory in quantum chemistry, this metric leverages the eigenvalue spectrum to quantify structural complexity \cite{gutman2001energy}.
    \item \textit{Laplacian Graph Energy (LGE):} Also introduced by Gutman \cite{gutman2006laplacian}, LGE derives from the Laplacian matrix eigenvalues and exhibits sensitivity to degree distribution and connectivity patterns. Both GE and LGE have demonstrated predictive power for system resilience and tipping-point identification \cite{edwards2024resilience}.
\end{itemize}

\textit{Structural Metrics:}
\begin{itemize}
    \item \textit{Cyclomatic Complexity:} Originally developed by McCabe \cite{mccabe1976complexity} for software control flow analysis, Cyclomatic Complexity counts the number of linearly independent paths through a graph structure. For a graph $G$ with $n$ nodes, $e$ edges, and $p$ connected components: $CC(G) = e - n + 2p$. McCabe and Butler argued that complexity assessments should be conducted before implementation to understand underlying complexity \cite{mccabe1989design}. While this metric does not satisfy all of Weyuker's criteria for complexity measures \cite{weyuker1988evaluating}, it has been related to cost and effort in software development projects and is adapted here for structural complexity assessment.
    \item \textit{Density:} Network density measures overall connectedness of a structure, computed as $D = \frac{e}{\frac{n \cdot (n-1)}{2}}$ for undirected graphs, where $e$ is edges and $n$ is nodes. While density does not measure complexity directly, it has been used to assess attributes similar to system complexity in real networks \cite{leskovec2005graphs, lei2019improved}. Higher density may signal increased verification effort and potential rework cycles in change management.
    \item \textit{Density Delta:} The difference between actual network density and the minimum possible density for a connected network of the same size. A tree or star topology represents minimum density ($d_{min} = \frac{2}{n}$) \cite{6549917}; Density Delta captures excess connectivity beyond this baseline \cite{vierlboeck2023re}.
    \item \textit{Absolute Density:} A size-adjusted density measure originally developed for social networks \cite{scott1988social} that accounts for network circumference, radius, and diameter, enabling comparisons across networks of different scales. This addresses the limitation that standard density naturally declines as networks grow \cite{lei2019improved}.
\end{itemize}

\textit{Composite Metrics:}
\begin{itemize}
    \item \textit{Load} ($L$): Defined as the total number of loops in a given network structure. Loops represent circular connections that can be problematic for satisfying requirements and managing changes, as modifications may propagate through circular dependencies. Due to its relevance to validation/verification processes and change management, this loop count is denoted as Load to distinguish it from components of other metrics \cite{vierlboeck2023re}:
    \begin{equation}
    L = |l_i|
    \end{equation}
    where $l_i$ represents the loops in the network. Integration Load computes this value for each integration task.
\end{itemize}

Table~\ref{tab:metrics_summary} summarizes the metrics evaluated across both analysis levels.

\begin{table}[H]
\centering
\caption{Summary of Complexity Metrics Evaluated.}
\label{tab:metrics_summary}
\begin{tabular}{lll}
\toprule
\textbf{Level} & \textbf{Metric} & \textbf{Type} \\
\midrule
Molecule & Total Cyclomatic Complexity & Structural \\
Molecule & Average Cyclomatic Complexity & Structural \\
Molecule & Average GE & Spectral \\
Molecule & Average LGE & Spectral \\
Molecule & Average Density & Structural \\
Molecule & Average Absolute Density & Structural \\
\midrule
Integration & Integration GE & Spectral \\
Integration & Integration LGE & Spectral \\
Integration & Integration Density & Structural \\
Integration & Integration Absolute Density & Structural \\
Integration & Integration Density Delta & Structural \\
Integration & Integration Load & Composite \\
\bottomrule
\end{tabular}
\end{table}

\subsection{Experimental Design: Molecular Integration Case Study}

The goal of this experiment was to establish a connection between the effort or time it took for a participant to understand the system and the complexity of the system as measured by a structural complexity metric. The hypothesis was not simply that more complex systems take longer---which would be tautological---but rather that \textit{specific spectral metrics} derived from the eigenvalue spectrum of graph representations capture the dimension of structural complexity that predicts integration effort. This is a testable claim because multiple plausible metrics exist, and the experiment discriminated among them: as the results demonstrated, some metrics (GE, LGE) strongly predicted effort while others (density, normalized Laplacian variants) did not, despite all being computable from the same graph representation. Task completion time---measured from task initiation to successful assembly---served as the primary dependent variable, capturing the effort dimension of integration complexity.

\begin{figure}[H]
    \centering
    \includegraphics[width=0.85\linewidth]{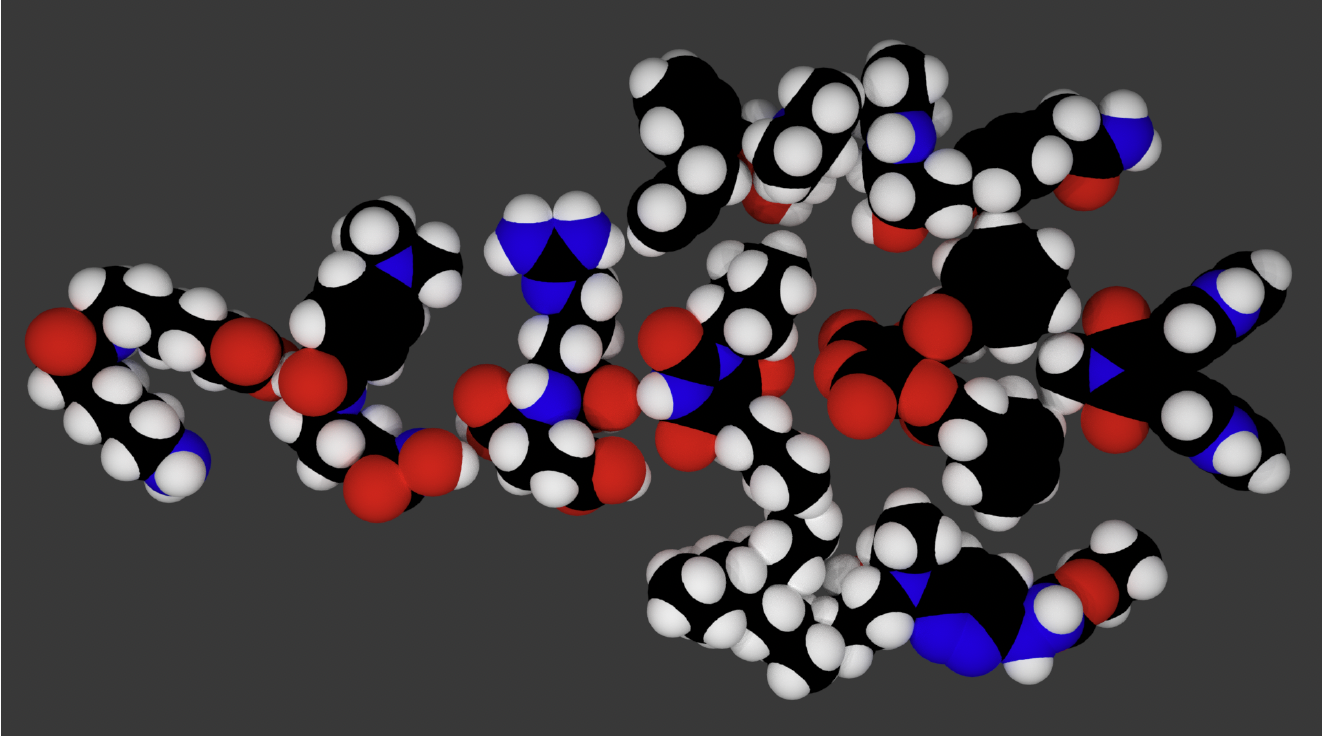}
    \caption{Example of completed object assembly. Colored spheres represent different atom types following standard CPK conventions.}
    \label{fig:moleculeassembly}
\end{figure}

Participants were asked to manipulate and integrate a series of objects that represented hydrocarbon molecule models to replicate an assembly of such objects in a tridimensional virtual environment (set up with the software Blender)---an example system adapted from \cite{pugliese2018development} is represented in Figure~\ref{fig:moleculeassembly}. The use of chemical models also removed any possible bias from preexisting knowledge of the system, such as native language or expertise in a specific field. Molecular structures were selected over other possible graph-based tasks (e.g., network routing, circuit assembly, abstract node-link diagrams) for several reasons: they provide precisely defined, repeatable graph topologies with known ground truth; they permit physical manipulation in three-dimensional space, engaging the spatial reasoning processes relevant to system integration; they naturally instantiate all three complexity contributions (component, interface, and topological) through atom types, bond types, and molecular topology; and they require no domain expertise from participants, ensuring that performance differences reflect structural complexity rather than prior knowledge \cite{pugliese2018development}.

Since the metrics under study were complexity metrics built to react to component complexity ($\alpha_i$), interface complexity ($\beta_{ij}$), and topological complexity as described in Section~\ref{subsec:metrics-framework}, all three contributions had to be addressed by the experiment. This was achieved by using objects whose complexity could be measured through the same metrics, applying the interface complexity model from Section~\ref{subsec:metrics-framework}, and measuring topological complexity through the matrix representation of the assembly.

Twenty-three participants were recruited from Stevens Institute of Technology. Participants had engineering or technical backgrounds but no prior experience with molecular modeling or chemistry, ensuring that domain expertise did not confound task performance. No demographic data beyond institutional affiliation were collected, representing a limitation for generalizability assessment. Each participant completed ten tasks administered in a randomized order. The tasks were divided into two groups of five tasks, with high and low expected complexity respectively, then randomized to avoid patterns learned in a specific task consistently impacting the completion time of subsequent tasks. Sample size adequacy was verified through Kolmogorov--Smirnov tests confirming normal distribution \cite{pugliese2018development}.

As established in Section~\ref{subsec:extending-metrics}, the structural isomorphism between molecular and requirement networks permits controlled manipulation of complexity parameters that would be confounded by semantic variation in natural language requirements.

It is important to delineate what this experimental design isolates and what it sets aside for future investigation. The design isolates \textit{structural complexity}---the topological properties of component-interface networks that affect integration effort---by controlling variables that would otherwise confound the analysis in requirements engineering contexts. Specifically, semantic content, linguistic ambiguity, stakeholder interpretation differences, and domain-specific knowledge are held constant through the use of domain-agnostic molecular structures, ensuring that observed performance differences are attributable to structural properties rather than these factors. This controlled isolation is a deliberate methodological strength: it establishes a clean baseline relationship between structural complexity \cite{sinha2014structural} and effort that would be impossible to identify in a full requirements engineering setting where multiple factors vary simultaneously. The correlations reported in Section~\ref{sec:results} therefore represent the \textit{structural component} of integration effort. In practice, the additional factors present in requirements work may interact with structural complexity, and characterizing these interactions is an important direction for future validation (Section~\ref{subsec:threats-to-validity}).

This study was reviewed and approved by the Stevens Institute of Technology Institutional Review Board. All participants provided informed consent prior to participation.

\section{Results}
\label{sec:results}

\subsection{Overall Case Study Results}
\label{subsec:overall-results}

\begin{figure}[H]
  \centering
  \includegraphics[width=1\textwidth]{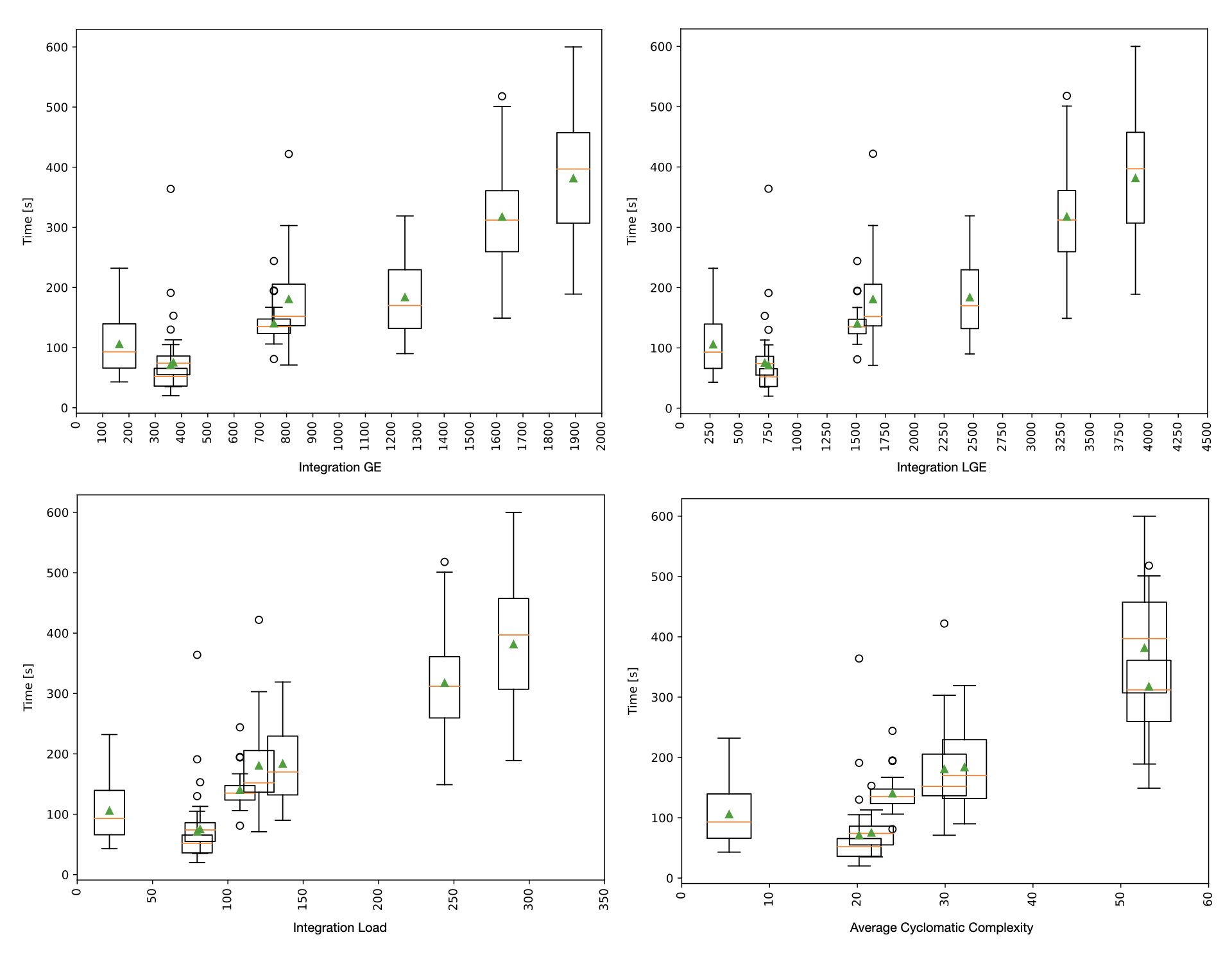}
  \caption{Box-plot representation of integration times versus metric values for the four top-performing metrics: Integration GE, Integration LGE, Integration Load, and Average Cyclomatic Complexity. The orange line indicates the median, the green triangle indicates the mean, and individual markers denote outliers. Box-plots for all remaining metrics are provided in the Appendix (Figure~\ref{fig:boxplot-appendix}).}
  \label{fig:boxplot}
\end{figure}

The results of the experiment are presented in Figure~\ref{fig:boxplot}.
Each plot represents the distribution of integration times, the horizontal axis being the value of the specific metric for each task.

A regression analysis of the data was performed using both a linear and a quadratic polynomial.
The application of such regression models assumes that the samples are independent and that the errors in the approximation are normally distributed.

The independence assumption was not strictly satisfied, since the same participant performed multiple tasks: individual ability affected more than one observation, and learning or fatigue effects may have carried across tasks. To address this limitation directly, Section~\ref{subsec:mixed-effects} presents a linear mixed-effects analysis with participant as a random intercept, which accounts for within-subject correlation and provides more conservative estimates of the complexity--effort relationship. Task order was randomized per participant and preceded by an untimed tutorial session \cite{pugliese2018development}, distributing any residual ordering effects across conditions rather than confounding them with complexity levels. Individual task sequences were not recorded in the original experimental protocol, precluding inclusion of task order as a covariate; this is addressed as a limitation in Section~\ref{subsec:limitations} and identified as a target for future experimental designs.

The second assumption, that the residuals were normally distributed, was verified.
The residuals had an almost normal distribution, with the quadratic model being closer than the linear one.
Given that the low end of these plots is not well approximated with a straight line, probably an exponential distribution would be a better approximation of the residuals.
This is expected given the high variance in human performance data, where a single-variable model cannot capture the full range of behavior.

Table~\ref{tab:statistical-analysis} presents the statistical results of the linear and quadratic polynomial regressions.
For all metrics and for both regression models, the \emph{p}-value is below the cut-off value of 0.05, meaning that there is strong evidence against the null hypothesis that the two variables in the model are not connected. This result indicates that both the linear and quadratic models are significant.

\begin{table}[H]
\caption{Statistical parameters for the linear and quadratic polynomial approximations: $r^2$ coefficient for the regression, coefficients of the polynomial $p(x) = \beta_0 + \beta_1 x + \beta_2 x^2$, and respective \emph{p}-values. The \emph{p}-values of the coefficients were evaluated through a \emph{t}-test, and values greater than the cutoff value of 0.05 are reported in bold.}
\label{tab:statistical-analysis}
\centering
\begin{tabular}{lrrrrrrr}
\toprule
\multirow{2}{*}{\textbf{Metric}} & \multirow{2}{*}{\boldmath{$r^2$}} & \multicolumn{2}{c}{\boldmath{$\beta_0$}} & \multicolumn{2}{c}{\boldmath{$\beta_1$}} & \multicolumn{2}{c}{\boldmath{$\beta_2$}} \\
\cmidrule(lr){3-4} \cmidrule(lr){5-6} \cmidrule(lr){7-8}
 &  & \textbf{Value} & \textbf{\emph{p}-Val} & \textbf{Value} & \textbf{\emph{p}-Val} & \textbf{Value} & \textbf{\emph{p}-Val} \\
\midrule
GE & 0.61 & 46.84 & 0.00 & 0.20 & 0.00 &  &  \\
 & 0.62 & 75.12 & 0.00 & 0.08 & \textbf{0.07} & 0.00 & 0.00 \\
\midrule
LGE & 0.61 & 49.80 & 0.00 & 0.08 & 0.00 &  &  \\
 & 0.61 & 63.71 & 0.00 & 0.05 & 0.00 & 0.00 & \textbf{0.14} \\
\midrule
NLGE & 0.15 & 42.78 & \textbf{0.07} & 3.76 & 0.00 &  &  \\
 & 0.19 & $-$107.81 & 0.03 & 13.10 & 0.00 & $-$0.13 & 0.00 \\
\midrule
NC & 0.36 & $-$304.54 & 0.00 & 18.90 & 0.00 &  &  \\
 & 0.37 & $-$24.23 & \textbf{0.93} & $-$4.57 & \textbf{0.84} & 0.48 & \textbf{0.29} \\
\midrule
LNC & 0.42 & $-$393.11 & 0.00 & 19.28 & 0.00 &  &  \\
 & 0.47 & 1510.04 & 0.00 & $-$117.22 & 0.00 & 2.39 & 0.00 \\
\midrule
NLNC & 0.18 & $-$169.81 & 0.00 & 15.05 & 0.00 &  &  \\
 & 0.20 & $-$711.90 & 0.01 & 65.26 & 0.01 & $-$1.13 & 0.04 \\
\midrule
GEn & 0.23 & $-$336.04 & 0.00 & 352.38 & 0.00 &  &  \\
 & 0.24 & 403.50 & \textbf{0.34} & $-$733.13 & \textbf{0.23} & 390.94 & \textbf{0.08} \\
\midrule
LGEn & 0.26 & $-$275.81 & 0.00 & 126.68 & 0.00 &  &  \\
 & 0.26 & 149.77 & \textbf{0.66} & $-$115.84 & \textbf{0.55} & 33.88 & \textbf{0.21} \\
\midrule
NLGEn & 0.04 & $-$129.76 & \textbf{0.21} & 84.70 & 0.00 &  &  \\
 & 0.16 & $-$5439.48 & 0.00 & 3133.20 & 0.00 & $-$433.49 & 0.00 \\
\midrule
NCn & 0.38 & $-$139.58 & 0.00 & 701.41 & 0.00 &  &  \\
 & 0.42 & 292.34 & 0.01 & $-$1431.02 & 0.01 & 2453.78 & 0.00 \\
\midrule
LNCn & 0.14 & 52.42 & 0.02 & 78.77 & 0.00 &  &  \\
 & 0.15 & $-$34.13 & \textbf{0.58} & 207.66 & 0.02 & $-$40.97 & \textbf{0.13} \\
\midrule
NLNCn & 0.02 & 683.91 & 0.00 & $-$497.75 & 0.03 &  &  \\
 & 0.15 & $-$41{,}625.84 & 0.00 & 83{,}834.22 & 0.00 & $-$41{,}966.46 & 0.00 \\
\bottomrule
\end{tabular}
\end{table}

The coefficient of determination, $r^2$, represents how much of the variance in the data can be captured by the model. GE and LGE both have $r^2 > 60\%$, NC, LNC, and NCn have $30\% < r^2 < 50\%$, while the rest of the metrics are below 30\%.

The coefficients of both the linear and quadratic models are reported as well in Table~\ref{tab:statistical-analysis}, together with their respective p-values. For most of the metrics, the quadratic coefficients are not significant, meaning that a linear representation is more appropriate.

The linear relationship in the case of GE is in disagreement with the findings presented by Sinha \cite{sinha2016empirical} regarding a superlinear relationship between integration time and Graph Energy. This discrepancy is not problematic given that the introduction of component and interface complexity weighting in the metrics leads to a different experimental setup than that used by Sinha \cite{sinha2016empirical}.

Metrics based on the normalized Laplacian matrix (NLGE, NLNC, NLGEn, NLNCn) exhibit an inverted relationship with complexity: their values decrease as the number of components increases. As noted by Pugliese \cite{pugliese2018development}, this inversion arises from the interaction between the normalized Laplacian representation and the $\gamma = 1/n$ coefficient. Because this behavior is opposite to the expected monotonic increase, we discarded the normalized Laplacian as a viable matrix representation for complexity measurement purposes. A detailed analysis of this phenomenon across random graph families is provided in \cite{pugliese2018development}.

The domain-agnostic nature of this experimental design strengthens the validity and broader applicability of these findings. By using molecular structures rather than domain-specific artifacts, the experiment eliminates confounding factors such as domain expertise, linguistic interpretation, and semantic ambiguity. The structural isomorphism between molecular graphs and requirement networks (established in Section~\ref{subsec:extending-metrics}) suggests that these results generalize to any domain where complexity can be represented through graph structures, including requirements engineering, system architecture analysis, and software design.

\subsection{Requirement-Specific Analysis Results}

Building upon the spectral metric analysis presented above, this section extends the investigation to metrics computed at both the molecule level (individual components) and the integration level (combined assemblies). This dual-level analysis, combined with additional structural metrics including Cyclomatic Complexity, Density, and Integration Load, provides insight into which measurements offer the strongest predictive power for effort estimation.

\subsubsection{Molecule-Level Metrics}
\label{subsec:molecule-metrics}

Table~\ref{tab:molecule_metrics} presents correlation coefficients between molecule-level complexity metrics and task completion time.

\begin{table}[H]
\centering
\caption{Molecule-level metric correlations with task completion time.}
\label{tab:molecule_metrics}
\begin{tabular}{lcc}
\toprule
\textbf{Metric} & \textbf{Correlation (r)} & \textbf{95\% CI} \\
\midrule
Total Cyclomatic Complexity & 0.8919 & [0.504, 0.9804] \\
Average Cyclomatic Complexity & 0.9125 & [0.5822, 0.9843] \\
Average Graph Energy & 0.9420 & [0.7059, 0.9897] \\
Average Laplacian Graph Energy & 0.9426 & [0.7086, 0.9898] \\
\rowcolor{gray!15} Average Density & $-$0.4163 & [$-$0.8667, 0.4081] \\
\rowcolor{gray!15} Average Absolute Density & $-$0.3446 & [$-$0.8443, 0.4756] \\
\bottomrule
\end{tabular}

\noindent{\footnotesize{\textcolor{gray}{Gray rows: not statistically significant (CI spans zero).}}}
\end{table}

The spectral metrics (Graph Energy and Laplacian Graph Energy) demonstrate strong positive correlations with task completion time ($r > 0.94$), with confidence intervals bounded well above zero. Similarly, Cyclomatic Complexity exhibits strong predictive validity at both total ($r = 0.8919$) and averaged ($r = 0.9125$) calculations. These results indicate that the complexity of individual components, as captured by spectral and structural metrics, is strongly indicative of the effort required to work with and integrate those components.

In contrast, density-based metrics show weak negative correlations that fail to achieve statistical significance, with confidence intervals spanning zero (Table~\ref{tab:molecule_metrics}, gray rows). This result is not marginal: the strongest density correlation ($r = -0.47$ for Integration Absolute Density) remains statistically indistinguishable from zero at the 95\% level. Connection density alone, without consideration of topological distribution, does not capture the aspects of complexity that affect human performance on integration tasks. This finding carries practical weight: density metrics are among the simplest and most commonly reported network statistics, yet they provide no predictive value for effort estimation in this context. Two graphs with identical density can exhibit markedly different eigenvalue spectra---and, as the data show, markedly different integration times. Practitioners should therefore treat density as a descriptive network statistic rather than a predictor of development effort that can be used to drive decisions that otherwise influence systems development.

\subsubsection{Integration-Level Metrics}
\label{subsec:integration-metrics}

Table~\ref{tab:integration_metrics} presents correlation coefficients for metrics computed at the integration level, capturing complexity when combining components.

\begin{table}[H]
\centering
\caption{Integration-level metric correlations with task completion time.}
\label{tab:integration_metrics}
\begin{tabular}{lcc}
\toprule
\textbf{Metric} & \textbf{Correlation (r)} & \textbf{95\% CI} \\
\midrule
Integration GE & 0.9545 & [0.7631, 0.992] \\
Integration LGE & 0.9572 & [0.7758, 0.9925] \\
Integration Load & 0.9546 & [0.7636, 0.992] \\
\rowcolor{gray!15} Integration Density & $-$0.3627 & [$-$0.8501, 0.4594] \\
\rowcolor{gray!15} Integration Absolute Density & $-$0.4720 & [$-$0.883, 0.3486] \\
\rowcolor{gray!15} Integration Density Delta & 0.3626 & [$-$0.4595, 0.8501] \\
\bottomrule
\end{tabular}

\noindent{\footnotesize{\textcolor{gray}{Gray rows: not statistically significant (CI spans zero).}}}
\end{table}

Integration-level spectral metrics exhibit even stronger correlations than their molecule-level counterparts. Integration LGE achieves the highest observed correlation ($r = 0.9572$), followed closely by Integration Load ($r = 0.9546$) and Integration GE ($r = 0.9545$). These results demonstrate that the complexity emerging from the combination of components and captured through spectral analysis of the integrated assembly serves as a robust predictor of integration effort.

The Integration Load metric, which combines component complexity ($\alpha_i$) and interface complexity ($\beta_{ij} = \sqrt{\alpha_i \alpha_j}$), performs comparably to the spectral metrics. This convergence across theoretically distinct approaches strengthens confidence that the underlying construct---structural complexity affecting integration effort---is being validly captured through multiple measurement pathways.

\subsubsection{Effectiveness of Spectral Metrics Versus Density}

The pattern across both analysis levels is clear: spectral metrics (GE, LGE) and structural metrics (Cyclomatic Complexity, Integration Load) demonstrate strong predictive validity, while density-based metrics consistently fail to correlate with task performance. This differential performance carries important implications for metric selection in practice.

Density measures the ratio of actual connections to possible connections but provides no information about how those connections are distributed topologically. A highly connected cluster and a uniformly distributed network may have identical density yet present markedly different integration challenges. The spectral metrics, by contrast, capture these topological distinctions through eigenvalue decomposition of the adjacency and Laplacian matrices. The eigenvalue spectrum encodes information about path lengths, clustering patterns, and structural regularity that proves cognitively relevant to integration tasks.

This finding refines earlier theoretical work proposing multiple candidate metrics for complexity assessment. While density metrics have intuitive appeal and appear in various network analysis frameworks \cite{wu2010natural, sinha2014structural}, they do not capture the aspects of structural complexity that manifest as increased effort during integration activities. Practitioners seeking to assess requirements complexity should prioritize spectral metrics over density-based measures.

The differential performance among spectral metrics themselves warrants a theoretical interpretation. Pugliese's generalized spectral complexity formula (Equation~(\ref{eq:generalized-complexity}), \cite{pugliese2019developing}) reveals that Graph Energy and Natural Connectivity share the same eigenvalue inputs but differ in their aggregation functions: GE applies linear aggregation ($f(x) = x$, $g(y) = |y|$), whereas NC applies exponential-then-logarithmic aggregation ($f(x) = \ln(x)$, $g(y) = e^y$). This distinction has a concrete mathematical consequence. In the NC formulation, the exponential function amplifies the largest eigenvalue disproportionately, causing the metric to be dominated by the spectral radius---effectively summarizing the graph by its single most prominent structural feature. GE, by contrast, weights all eigenvalues proportionally through linear summation, capturing the full distribution of topological properties including path lengths, branching patterns, and clustering across the entire structure. For integration tasks, where a participant must comprehend and assemble all components and interfaces, the cumulative contribution of every structural feature is cognitively relevant. A structure with many moderately complex branches imposes integration effort that GE reflects through its distributed aggregation but that NC underweights relative to any single dominant substructure. Additionally, NC was originally developed as a network robustness measure \cite{wu2010natural}---quantifying resistance to node or edge removal---rather than as a complexity measure per se. Robustness and cognitive integration effort are distinct constructs; a robust network may or may not be cognitively demanding to assemble. The empirical results ($r^2 > 0.60$ for GE versus $r^2 = 0.36$ for NC) are consistent with this theoretical analysis: metrics that capture distributed structural properties outperform those dominated by a single spectral feature when the prediction target is human integration effort.

\subsubsection{Mixed-Effects Analysis}
\label{subsec:mixed-effects}

The regression results presented in Sections~\ref{subsec:overall-results}, \ref{subsec:molecule-metrics} and \ref{subsec:integration-metrics} treated each observation as independent. However, because the same twenty-three participants each completed multiple tasks, observations within a participant were correlated. To account for this within-subject dependence, linear mixed-effects models were fitted via restricted maximum likelihood (REML) with participant as a random intercept and each complexity metric as the sole fixed-effect predictor. This analysis was only possible to be conducted for eight of the 10 tasks since retrospect calculations were not possible anymore, years after the experiment was conducted. This still yielded $N = 184$ for the number of observations.

Table~\ref{tab:mixed_effects} summarizes the results for the seven key metrics.

\begin{table}[H]
\centering
\caption{Linear mixed-effects model results with participant as random intercept. $\beta$ is the fixed-effect slope, $R^2_m$ is the marginal $R^2$ (variance explained by the metric alone), $R^2_c$ is the conditional $R^2$ (metric plus participant effects), and ICC is the intraclass correlation coefficient representing the proportion of residual variance attributable to between-participant differences.}
\label{tab:mixed_effects}
\begin{tabular}{lccccccc}
\toprule
\textbf{Metric} & \boldmath{$\beta$} & \textbf{SE} & \boldmath{$t$} & \boldmath{$p$} & \boldmath{$R^2_m$} & \boldmath{$R^2_c$} & \textbf{ICC} \\
\midrule
Int. LGE    & 0.0835 & 0.0044 & 18.89 & $<$0.001 & 0.625 & 0.678 & 0.141 \\
Int. GE     & 0.1713 & 0.0092 & 18.71 & $<$0.001 & 0.621 & 0.674 & 0.138 \\
Int. Load   & 1.2114 & 0.0647 & 18.72 & $<$0.001 & 0.621 & 0.674 & 0.138 \\
Avg. GE     & 0.4818 & 0.0268 & 17.97 & $<$0.001 & 0.605 & 0.655 & 0.127 \\
Avg. LGE    & 0.2359 & 0.0131 & 18.00 & $<$0.001 & 0.606 & 0.656 & 0.127 \\
Avg. Cyclomatic & 6.3155 & 0.3847 & 16.41 & $<$0.001 & 0.568 & 0.612 & 0.103 \\
Total Cyclomatic & 0.5048 & 0.0326 & 15.47 & $<$0.001 & 0.542 & 0.583 & 0.090 \\
\bottomrule
\end{tabular}
\end{table}

All seven metrics remain highly significant ($p < 0.001$) after accounting for between-participant variance. The intraclass correlation coefficient (ICC $= 0.14$ for Integration LGE) indicates that approximately 14\% of the unexplained variance in completion time is attributable to stable individual differences among participants (e.g., spatial reasoning ability, working memory capacity), while the remaining 86\% reflects within-participant, task-level variation. This confirms that individual differences exist but are moderate relative to the effect of task complexity.

The marginal $R^2$ values (0.54--0.63), computed following the method of Nakagawa and Schielzeth \cite{nakagawa2013}, represent the variance in individual completion times explained by the complexity metric alone, providing a more conservative estimate than the $r^2 > 0.60$ obtained from task-mean regressions in Table~\ref{tab:statistical-analysis}. The difference is expected: aggregating to task means removes within-task variability, inflating the apparent fit. The individual-level marginal $R^2$ is the more appropriate measure for characterizing predictive power at the observation level.

The conditional $R^2$ values (0.58--0.68) show that adding participant-level random intercepts explains an additional 4--5\% of variance beyond the metric alone. This modest increment further confirms that task complexity, not individual differences, is the dominant driver of completion time variation.

The ranking of metrics is preserved from the correlation analysis: Integration LGE achieves the highest marginal $R^2$ (0.625), followed by Integration GE (0.621) and Integration Load (0.621), with molecule-level metrics and Cyclomatic Complexity following. This consistency across analytical methods strengthens confidence in the robustness of the complexity--effort relationship.

\subsection{Application to Requirements Complexity}

The molecular integration results, while not conducted on natural language requirements, suggest potential implications for requirements engineering practice that warrant future direct validation through the structural isomorphism and NLP extraction methodology established in Section~\ref{subsec:extending-metrics}. This section discusses how the validated metrics can be applied to predict integration challenges in requirements engineering contexts.

\subsubsection{From Molecular Models to Requirements Structures}

The strong correlations observed between spectral complexity metrics and integration effort in molecular tasks suggest effects that may be expected when working with requirements of varying structural complexity, pending direct validation in RE contexts. As demonstrated in prior work \cite{vierlboeck2025natural}, requirements specifications can be decomposed into structural networks that exhibit topological properties analogous to those of molecular assemblies. Requirements that form dense clusters of interdependencies, similar to ring-based molecular structures, can be expected to present greater integration challenges than those with simpler, more hierarchical arrangements.

These results suggest that the same metrics that predict molecular integration effort may apply to requirement structures extracted through NLP methods, though direct validation on requirements integration tasks is needed to confirm this transfer. Graph Energy and Laplacian Graph Energy, computed on requirement networks, can serve as leading indicators of the effort required to integrate, implement, and maintain those requirements. This connection was established theoretically in previous research \cite{vierlboeck2025natural, SysCon2022}; the present study provides the empirical validation.

\subsubsection{Predicting Integration Challenges}

The practical application of these findings proceeds as follows: requirement specifications are processed through the NLP extraction methodology \cite{vierlboeck2025natural}, yielding structural networks representing hierarchy, dependencies, and entity relationships. From these extracted structures, spectral complexity metrics (GE, LGE) and structural metrics (Cyclomatic Complexity, Integration Load) can be computed. If these results transfer to requirements contexts, higher metric values may indicate requirements sets that demand greater integration effort.

This predictive capability addresses a critical gap in requirements engineering practice. Currently, complexity assessments rely primarily on expert judgment or surface-level text metrics such as requirement count and word length. These approaches do not capture the structural interdependencies that drive integration difficulty. Spectral metrics, by contrast, quantify the topological properties that emerge from requirement relationships. These properties are the same that affect how engineers comprehend, trace, and implement system specifications.

\subsubsection{Domain Agnosticism and Broader Applicability}

A significant advantage of the spectral complexity approach is its domain-agnostic nature. The metrics operate on graph representations rather than domain-specific content, enabling application across diverse engineering contexts. Whether analyzing software requirements, aerospace system specifications, or organizational process models, the same spectral measures apply once the underlying structure has been extracted.

This agnosticism was deliberately built into the experimental design by using molecular structures rather than natural language requirements. Participants had no domain expertise in chemistry, ensuring that task performance reflected structural complexity rather than prior knowledge. The strong correlations observed across participants with varying backgrounds support the generalizability of these metrics to other domains where complexity manifests through structural interdependencies.

These results provide preliminary evidence that complements prior research on requirements complexity \cite{vierlboeck2025natural}, which established the NLP extraction methodology and proposed spectral metrics as candidates for complexity assessment. That work demonstrated that complexity could be measured from requirements; the present study demonstrates that measured complexity matters---it predicts the effort required to integrate complex structures. Together, these findings establish a foundation for proactive complexity management in requirements engineering, enabling practitioners to identify and address problematic requirement structures before they cascade into downstream development challenges.

\section{Discussion}
\label{sec:discussion}

\subsection{Implications for Systems Development}

The experimental results presented in Section~\ref{sec:results} establish a clear connection between structural complexity metrics and the effort required to complete integration tasks. This connection carries significant implications for systems development practice, particularly regarding development time, cost, and the management of system quality attributes.

\subsubsection{Complexity, Development Time, and Cost}

The correlations found in this study demonstrate that higher levels of complexity, as measured by Graph Energy, Laplacian Graph Energy, and Integration Load, correspond to higher effort and more time required from humans working with specific requirement configurations. These findings suggest that a set of requirements with higher complexity may have a longer development cycle, though direct validation in RE contexts is needed to confirm this transfer. Since these factors are directly tied to project cost and schedule, complexity metrics computed on requirements can serve as leading indicators of downstream development challenges \cite{bhatnager2025measuring}. When requirement specifications exhibit high complexity values, project managers can anticipate increased integration effort and allocate resources accordingly \cite{BASHIR2004195, Bashir01091999}.

\subsubsection{Impact on System Quality Attributes}

Beyond schedule and cost, structural complexity influences system quality attributes---often referred to as the ``ilities'' (maintainability, reliability, testability, etc.). Requirements that form dense clusters of interdependencies create cascading effects throughout the system life cycle. High complexity in requirement specifications correlates with:

\begin{itemize}
    \item Difficult change implementation and reduced maintainability: Complex interdependencies make it difficult to modify individual requirements without affecting others, increasing the effort and risk associated with system updates or engineering changes.
    \item Increased testing effort: Requirements with high Cyclomatic Complexity and integration load require more extensive verification activities, as the number of logical paths and interaction combinations grows.
    \item Decreased reliability: The correlation between complexity and human error suggests that complex requirement structures are more prone to misinterpretation during implementation, potentially introducing defects that affect system reliability.
\end{itemize}

These effects compound over the system life cycle. Complexity introduced at the requirements stage propagates through architecture, design, and implementation, amplifying its impact on development effort and system quality.

\subsubsection{Resource Allocation and Risk Mitigation}

The identification and monitoring of high-complexity requirement specifications enable informed resource allocation and risk mitigation strategies. Using the NLP-based extraction methodology, the detection algorithm can identify requirements that merit scrutiny, pointing to areas of interest that could be assessed by a human in the loop. By computing spectral complexity metrics during requirements development, project teams can:

\begin{itemize}
    \item Identify complexity hotspots: Requirements or requirement clusters exhibiting unusually high Graph Energy or Integration Load values warrant additional scrutiny and potentially decomposition into simpler structures.
    \item Inform staffing decisions: Integration activities involving high-complexity requirement sets may require more experienced engineers or larger teams to manage the increased cognitive load.
    \item Prioritize risk mitigation: Requirements with high structural complexity represent elevated risk areas where additional reviews, prototyping, or early integration testing may be warranted.
    \item Support trade-off decisions: When design alternatives exist, complexity metrics provide quantitative input for selecting approaches that minimize integration challenges.
    \item Monitor complexity trends: By tracking complexity metrics over time, teams can detect unnoticed increases or jumps in complexity---often referred to as complexity creep---before consequences become visible and even potentially irrevocable \cite{vierlboeck2025natural}.
    \item Benchmark against previous projects: Comparing complexity levels against other or previous projects enables identification of disproportionate complexity levels, allowing teams to flag specifications that deviate significantly from established baselines.
\end{itemize}

By monitoring and analyzing such metrics, projects can be better controlled, helping to reduce risk by providing a way to identify and address potential problems early in development.

This capability for early complexity assessment addresses a critical gap in current practice. While architectural complexity can be assessed once designs are sufficiently developed, requirements complexity has historically lacked equivalent quantification methods as shown above. The spectral metrics validated here enable complexity management to begin at the earliest development stages, when intervention costs are lowest and design flexibility is greatest.

\subsection{Integration of Metrics into Requirements Engineering}

The empirical validation presented in this study establishes spectral complexity metrics as viable predictors of integration effort. However, for these metrics to deliver practical value, they must be systematically integrated into requirements engineering workflows. This section proposes an integration framework, contrasts spectral approaches with traditional RE metrics, and discusses the practical implications for requirements management.

\subsubsection{A Framework for Metric Integration}

The integration of spectral complexity metrics into RE processes follows naturally from the NLP-based extraction methodology established in prior work \cite{vierlboeck2025natural, SysCon2022}. We propose a three-stage integration framework:

\textit{Stage 1: Structural Extraction.} Requirements specifications are processed through the NLP pipeline described in Section~\ref{subsec:extending-metrics} to extract the three structural layers \cite{vierlboeck2025natural}. This extraction achieves over 99 percent precision \cite{vierlboeck2023re}.

\textit{Stage 2: Complexity Quantification.} From the extracted layers, weighted adjacency matrices are constructed and spectral complexity metrics computed. Based on experimental results, Graph Energy, Laplacian Graph Energy, and Integration Load should be prioritized; density-based metrics should be de-prioritized.

\textit{Stage 3: Decision Support.} Computed complexity values inform engineering decisions throughout the development life cycle. High-complexity requirements can be flagged for review, decomposition, or resource allocation.

If validated for requirements contexts, this framework would enable complexity management at the earliest development stages, when intervention costs are lowest and design flexibility is greatest.

To illustrate, consider the Skyzer UAV landing gear specification analyzed in prior work \cite{vierlboeck2023re, blackburn2018skyzer}. In Stage~1, the NLP pipeline extracts 79 requirements and 246 unique entities, producing an entity-level adjacency matrix that reveals dense clusters of interdependent requirements---topologically resembling cyclic molecular structures such as 1,3,5-triazine (Figure~\ref{fig:requirementsanalogy}(b)). In Stage~2, spectral metrics are computed on this extracted network: Graph Energy and Laplacian Graph Energy quantify the overall structural complexity, while Integration Load captures the combined component and interface contributions. Based on the correlations validated in the present study, high metric values in specific requirement clusters would predict elevated integration effort for the corresponding subsystems. In Stage~3, systems engineers use these quantified complexity values to prioritize resources---for instance, directing additional integration testing toward the high-complexity landing gear actuation cluster while applying lighter oversight to structurally simpler subsystems. This targeted allocation is possible only because the spectral metrics distinguish topological complexity from simple requirement count: two subsystems with equal numbers of requirements may exhibit markedly different spectral complexity and, consequently, different integration risk and complexity profiles.

\subsubsection{Contrast with Traditional Requirements Engineering Metrics}

Traditional RE complexity metrics predominantly focus on syntactic and semantic properties of individual requirements, as discussed in Section~\ref{subsec:complexity-in-re}. Text-based approaches evaluate requirement length, readability, and linguistic ambiguity \cite{Berry2004, kamsties2001detecting}, while network-based approaches model requirements as graphs to analyze connectivity and centrality \cite{hein2022reasoning}. While these metrics provide useful insights into requirement quality, they do not capture the structural dependencies that drive integration difficulty. This study did not include direct empirical comparison between spectral metrics and these traditional approaches; however, the theoretical limitations of text-based metrics---specifically their inability to capture structural interdependencies---suggest that spectral metrics address a fundamentally different dimension of complexity. Empirical validation comparing predictive performance across metric types remains an important direction for future research.

Network-based approaches have emerged that model requirements as graphs, analyzing properties such as connectivity and centrality \cite{hein2022reasoning}. These methods offer valuable perspectives on interdependencies but typically do not leverage spectral graph properties, which limits their ability to quantify deeper structural complexities.

The spectral approach offers three distinct advantages over traditional metrics:

\begin{enumerate}
    \item Structural depth: Spectral metrics capture topological properties through eigenvalue decomposition of adjacency and Laplacian matrices. The eigenvalue spectrum encodes information about path lengths, clustering patterns, and structural regularity that text-based metrics cannot access. Two requirement sets with identical word counts and readability scores may exhibit vastly different spectral complexity values if their dependency structures differ.

    \item Empirical grounding: Unlike many traditional RE metrics, which are proposed based on intuition or analogy to software metrics, the spectral metrics validated in this study demonstrate direct correlation with human performance on integration tasks. This empirical foundation provides confidence that measured complexity reflects actual cognitive and effort demands.

    \item Early applicability: Spectral metrics can be computed as soon as requirements exist in textual form, before architectural decisions constrain design options. Traditional architectural complexity metrics \cite{sinha2016empirical} require sufficiently developed system representations that may not be available until later development phases.
\end{enumerate}

\subsubsection{Practical Usage and Implementation Considerations}

The practical deployment of spectral complexity metrics requires consideration of several factors. First, the NLP extraction pipeline must be configured appropriately for the domain and terminology of the requirements under analysis. While the syntactic parsing approach demonstrates broad applicability, domain-specific entity recognition may require calibration.

Second, complexity thresholds must be established contextually. The absolute values of Graph Energy or Integration Load depend on specification size and domain characteristics. Rather than universal thresholds, organizations should establish baseline complexity profiles for their typical specifications and flag deviations that exceed historical norms.

Third, complexity metrics should complement rather than replace human judgment. The metrics identify structural patterns warranting attention; engineers must then assess whether high complexity reflects genuine system needs or indicates opportunities for simplification. Such insights can point to requirements of interest that merit scrutiny and could be assessed by a human in the loop.

Finally, the approach can be extended to detect requirements containing potential inconsistencies by reviewing the results of the identified entities and checking them for consistency \cite{vierlboeck2025natural}. This capability provides secondary benefits beyond complexity measurement, improving overall requirements quality through automated structural analysis.

When integrated effectively, spectral complexity metrics transform requirements engineering from a predominantly qualitative discipline into one informed by quantitative structural analysis, which enables proactive complexity management that has long been available for system architectures but absent from requirements practice.

\subsection{Threats to Validity}
\label{subsec:threats-to-validity}

The validity of these findings can be assessed along three dimensions. Construct validity concerns whether the molecular integration task and the chosen metrics actually measure structural complexity and effort as intended. The structural isomorphism established in Section~\ref{subsec:extending-metrics} and the convergence of multiple independent metrics (GE, LGE, Cyclomatic Complexity, Integration Load) toward consistent predictions support construct validity; however, the proxy nature of the molecular task means that the construct being measured is structural integration effort, not requirements integration effort in full. Internal validity concerns whether the observed correlations reflect genuine complexity--effort relationships rather than confounds. The mixed-effects analysis (Section~\ref{subsec:mixed-effects}), task randomization, and domain-agnostic design mitigate the primary internal threats, though unrecorded task order remains a limitation (Section~\ref{subsec:limitations}). External validity concerns whether results generalize beyond the experimental setting. The molecular experiment deliberately controlled away semantic content, stakeholder factors, and domain knowledge to isolate structural effects. Whether the observed correlations hold, amplify, or attenuate when these factors are reintroduced in actual requirements engineering contexts remains the central open question requiring direct validation.

\subsubsection{Limitations of the Molecular Analogy}

The experimental design employed molecular integration tasks as a proxy for requirements integration based on the structural isomorphism described in Section~\ref{subsec:extending-metrics}. While this isomorphism is grounded in graph-theoretic equivalence, the analogy has inherent boundaries.

Molecular integration involves physical manipulation in a three-dimensional virtual environment, whereas requirements integration involves cognitive processing of semantic content, traceability establishment, and consistency verification. Requirements engineering presents distinct challenges---including subjective interpretation, linguistic ambiguity, and context-dependent \linebreak terminology---that constrain the generalizability of automated approaches across domains. Consequently, while the correlations observed are indicative of effects that can be expected in requirements work, the precise magnitude in actual RE contexts requires direct validation.

\subsubsection{Participant Variability and Sample Size}
\label{subsec:limitations}

The experiment involved twenty-three participants completing ten tasks each. While this design provided sufficient statistical power to detect strong correlations, individual differences in spatial reasoning, working memory capacity, and problem-solving strategies introduce variability. The mixed-effects analysis presented in Section~\ref{subsec:mixed-effects} directly addresses this concern: by modeling participant as a random intercept, between-subject variance is separated from task-level effects. The resulting ICC of 0.14 confirms that individual differences account for a meaningful but moderate proportion of unexplained variance, while task complexity remains the dominant predictor.

While task randomization was employed to mitigate systematic ordering effects, individual task sequences were not recorded in the original experimental protocol, precluding inclusion of task order as a covariate. The randomization distributes any learning or fatigue effects across conditions rather than eliminating them entirely. The experimental design included an untimed practice task with tutorial to reduce initial learning effects \cite{pugliese2018development}. Because tasks were divided into high- and low-complexity groups and then randomized per participant, any residual temporal drift (learning, fatigue) is orthogonal to the complexity conditions rather than confounded with them. Future studies should record individual task sequences and include task order as a covariate or employ counterbalanced designs to quantify these effects more precisely.

The adequacy of the sample size was assessed through Kolmogorov--Smirnov one-sample tests comparing each task with a normal distribution \cite{pugliese2018development}. Results indicated that all tasks except one were normally distributed at a significance level of 0.05, demonstrating that the sample of twenty-three participants was sufficiently large to be representative of the general population under the normal distribution assumption.

\subsubsection{Scaling to Large, Complex Systems}

The experimental tasks involved molecular structures of moderate complexity, considerably smaller than industrial-scale requirement specifications. Several scaling considerations emerge for larger systems.

First, while the NLP-based extraction achieved high precision and low error rates for the definition of requirement structures with current Natural Language Processing tools and libraries \cite{vierlboeck2023re}, heterogeneous requirement formats or specifications spanning multiple documents or specifications may present additional challenges.

Second, there is currently no standardized model to characterize the complexity of a system or network \cite{lei2019improved}, which complicates the establishment of universal thresholds for metric interpretation.

Third, some researchers have attempted to decouple their metrics from the size and scope of the network, but in doing so introduced other potential drawbacks \cite{lei2019improved}. The interpretation of absolute metric values becomes challenging at scale, suggesting that relative complexity comparisons within specification families may be more practical than absolute thresholds.

These considerations also limit comparability across systems. While similar systems may be compared with respect to their measured complexity, transferability across heterogeneous systems is not inherently guaranteed.

\subsubsection{Scope of Validated Metrics}

The present study validated correlations between complexity metrics and integration time as a proxy for effort. While determining causal relationships and correlations between different complexity metrics can be challenging, these metrics all contribute to the overall system and development complexity. Requirements introduce complexity to the process by increasing effort or load, and the metrics validated here capture one dimension of this relationship.

Additionally, while density-based metrics showed no predictive validity in this study (Section~\ref{subsec:molecule-metrics}), density appears in various network analysis frameworks \cite{sinha2014structural, wu2010natural} and may correlate with other aspects of system behavior not captured in the integration task paradigm.

\subsubsection{Comparison with Traditional RE Metrics}

This study validated spectral complexity metrics against integration effort but did not directly compare predictive performance against traditional RE metrics such as requirement count, word length, or readability scores. While Section~\ref{subsec:complexity-in-re} outlines the theoretical limitations of text-based approaches, empirical comparison would strengthen claims regarding the advantages of the spectral approach. Future work should include head-to-head comparisons between spectral metrics and traditional RE metrics on the same requirement sets to quantify the predictive improvement offered by structural analysis.

Despite these limitations, the consistent pattern of strong correlations across multiple spectral metrics provides a robust foundation for continued development of complexity-informed requirements engineering practices.

\section{Conclusions}
\label{sec:conclusion}

The inability to effectively characterize, quantify, and manage complexity remains a fundamental challenge in systems engineering. This research addressed that challenge by providing empirical validation that spectral complexity metrics predict integration effort, which bridges a critical gap between architectural complexity analysis, which has been extensively studied, and requirements complexity assessment, which has historically lacked equivalent quantification methods. The central finding is clear: complexity introduced at the requirements stage can be measured, and measured complexity predicts real effort.

\subsection{Summary of Contributions}

This research provides four primary contributions to the systems engineering body of knowledge.

\subsubsection{Isomorphic Experimental Methodology for Complexity Metric Validation}

By designing a controlled experiment using molecular integration tasks as structurally isomorphic proxies for requirements, this work established a reusable methodology for validating complexity metrics. The molecular analogy isolated structural complexity from confounding factors such as domain expertise, semantic ambiguity, and stakeholder familiarity, enabling rigorous empirical testing of metric--effort relationships that would be difficult to control in a direct requirements setting.

\subsubsection{Empirical Validation of Spectral Metrics as Effort Predictors}

Through controlled experimentation with twenty-three participants, we demonstrated that Graph Energy and Laplacian Graph Energy predicted integration effort with task-mean correlations exceeding 0.95, while structural metrics such as Cyclomatic Complexity achieved correlations above 0.89. A mixed-effects analysis accounting for between-participant variance confirmed these relationships at the individual observation level (marginal $R^2 = 0.63$ for Integration LGE, $p < 0.001$). Notably, density-based metrics showed no significant predictive validity, refining earlier theoretical work and providing actionable guidance for metric selection in practice.

\subsubsection{Preliminary Foundation for Extending Complexity Assessment to Requirements Engineering, Pending Direct Validation}

By establishing the structural isomorphism between molecular graphs and requirement networks, and by building upon NLP-based extraction methodologies that achieve over 99 percent precision in structural detection \cite{vierlboeck2023re}, this research enables complexity quantification at the earliest development stages, when intervention costs are lowest and design flexibility is greatest.

\subsubsection{A Framework for Proactive Complexity Management}

The three-stage integration framework (structural extraction, complexity quantification, decision support) transforms requirements engineering from a predominantly qualitative discipline into one informed by quantitative structural analysis. This enables practitioners to identify complexity hotspots, monitor complexity trends over time, and benchmark against previous projects before problematic patterns cascade into downstream development challenges.

Together, these contributions establish that while prior work demonstrated complexity can be measured from requirements, the present study demonstrates that measured complexity matters and predicts the effort required to integrate complex structures.

\subsection{Future Work}

Several avenues for future research emerge from this work.

\subsubsection{Integration with Large Language Models}

As LLMs increasingly enter requirements engineering workflows, validated complexity metrics become essential for assessing whether AI-generated specifications introduce problematic structural complexity. The spectral metrics validated in this study are well suited to serve as quality gates in such workflows. Consider a scenario in which an LLM drafts or refines a set of system requirements: the NLP pipeline established by Vierlboeck et al.\ \cite{vierlboeck2025natural} extracts the structural network from the generated text, and spectral metrics (GE, LGE) are computed on the resulting graph. Recent work has demonstrated the feasibility of such domain-specific pipelines, using language models to classify requirements and extract structural entities from natural language specifications \cite{tikayat2023agile}. If the integration-level complexity exceeds a threshold calibrated from historical project data, the specification is flagged for a critical human review before proceeding to design. Because the metrics are domain-agnostic and computable in seconds, this gating step introduces negligible overhead while providing an objective structural check that complements the semantic review a human engineer would perform. Future work should explore such automated pipelines that compute spectral metrics on LLM-generated requirements in real time, providing immediate feedback on structural quality before specifications are finalized. Longitudinal application would further enable trend monitoring across successive specification revisions, detecting complexity creep before its consequences propagate downstream.

\subsubsection{Machine Learning for Complexity Forecasting}

The correlations validated here suggest potential for predictive models that forecast integration effort based on early-stage complexity measurements. Supervised learning approaches trained on historical project data could enable more accurate schedule and resource estimation, while trend analysis over the project life cycle could detect complexity creep before consequences become visible and potentially irrevocable \cite{vierlboeck2025natural}.

\subsubsection{Direct Validation on Requirements Integration Tasks}

While the molecular integration experiment provides strong evidence through structural isomorphism, direct validation using natural language requirements would strengthen generalizability. Future studies should examine whether the observed correlations hold when participants work with actual requirement specifications across diverse domains.

\subsubsection{Scaling to Industrial Systems}

The experimental tasks involved structures of moderate complexity. Validation on industrial-scale specifications with thousands of requirements would address questions of computational scalability and metric interpretation at scale, potentially requiring hierarchical decomposition strategies or normalized comparison approaches.

\subsubsection{Comparative Validation Against Traditional Metrics}

Future studies should compare the predictive performance of spectral complexity metrics against traditional RE metrics such as requirement count, readability scores, and linguistic ambiguity measures. Such comparisons would quantify the added value of structural analysis over text-based approaches and identify contexts where each metric type proves to be most useful.

Ultimately, this research provides a foundation for treating requirements complexity with the same quantitative rigor that has long been applied to system architectures---enabling proactive complexity management at the inception of system development, where it can matter most.

\vspace{12pt}

\subsection*{Author Contributions}
Conceptualization, M.V. and A.P.; methodology, A.P. and M.V.; software, A.P.; validation, M.V., A.P. and R.R.N.; formal analysis, M.V. and A.P.; investigation, A.P.; resources, R.R.N.; data curation, A.P.; writing---original draft preparation, M.V. and A.P.; writing---review and editing, M.V., R.R.N., P.T.G. and R.S.N.B.; visualization, M.V. and A.P.; supervision, R.R.N.; project administration, M.V. and R.R.N. All authors have read and agreed to the published version of the manuscript.

\subsection*{Funding}
This research received no external funding.

\subsection*{Institutional Review Board Statement}
The study was approved by the Institutional Review Board of the Stevens Institute of Technology (protocol code 2018-025(N), 14 April 2018).

\subsection*{Informed Consent Statement}
Informed consent was obtained from all subjects involved in the study.

\subsection*{Data Availability Statement}
The experimental task completion time data and computed metric values supporting the findings of this study are openly available on Zenodo at \url{https://doi.org/10.5281/zenodo.19324374} under a Creative Commons Attribution 4.0 International (CC BY 4.0) license. The experimental protocol and metric derivations are documented in \cite{pugliese2018development}.

\subsection*{Conflicts of Interest}
The authors declare no conflicts of interest.

\subsection*{Abbreviations}
The following abbreviations are used in this manuscript:

\medskip
\noindent
\begin{tabular}{@{}ll}
AI   & Artificial Intelligence \\
GE   & Graph Energy \\
ICC  & Intraclass Correlation Coefficient \\
IRB  & Institutional Review Board \\
LGE  & Laplacian Graph Energy \\
LLM  & Large Language Model \\
NC   & Natural Connectivity \\
NLGE & Normalized Laplacian Graph Energy \\
NLP  & Natural Language Processing \\
OLS  & Ordinary Least Squares \\
RE   & Requirements Engineering \\
REML & Restricted Maximum Likelihood \\
\end{tabular}

\appendix
\section{Supplementary Figure}
\label{sec:appendix}

\begin{figure}[H]
  \centering
  \includegraphics[width=0.85\textwidth]{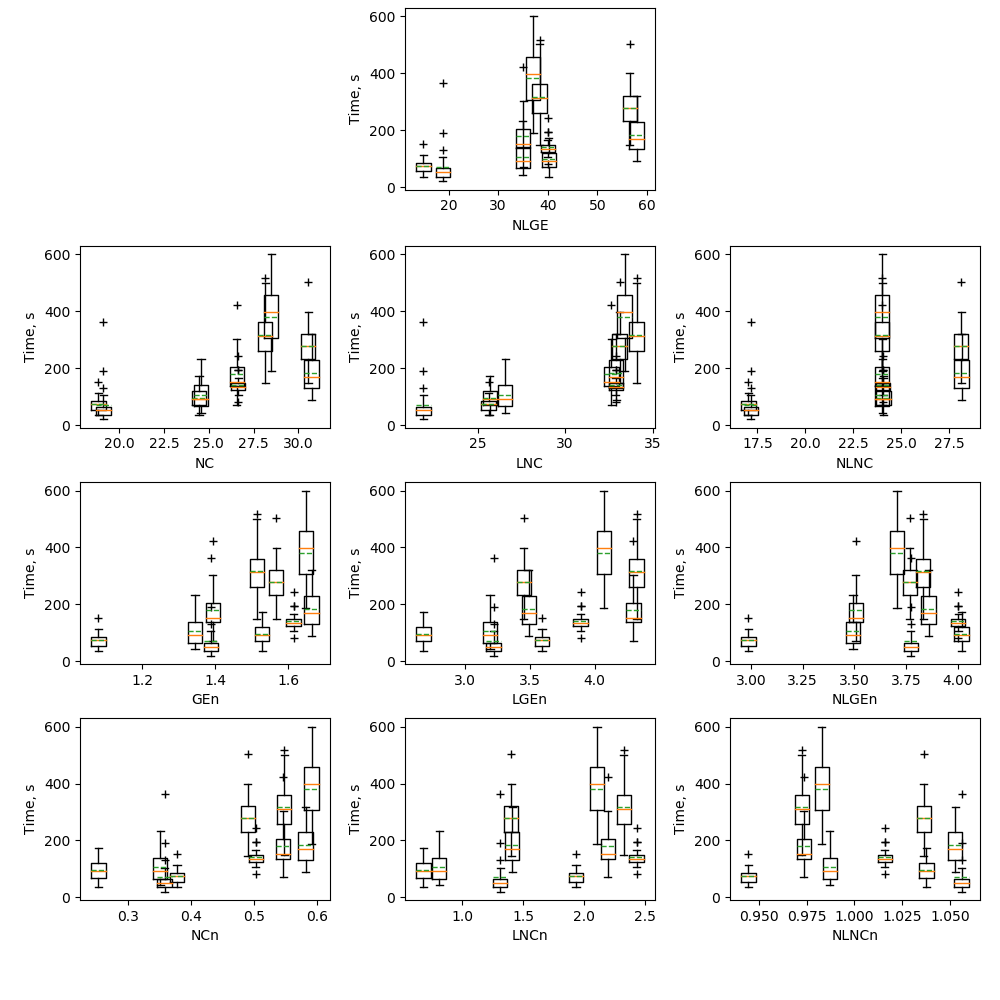}
  \caption{Box-plot representation of integration times versus metric values for the ten metrics not included in Figure~\ref{fig:boxplot}: NLGE, NC, LNC, NLNC, GEn, LGEn, NLGEn, NCn, LNCn, and NLNCn. The orange line indicates the median, the green triangle indicates the mean, and individual markers denote outliers.\label{fig:boxplot-appendix}}
\end{figure}

\bibliography{Reference_arxiv}

\end{document}